\documentclass[aps,prl,twocolumn,superscriptaddress,nopacs]{revtex4-1}

\usepackage{graphicx}
\usepackage{amssymb}
\usepackage{amsmath}
\usepackage{epstopdf}
\usepackage{mciteplus}
\usepackage{setspace}

\begin{document}

\title{Hiding in plain view: Colloidal self-assembly from polydisperse populations}

\author{Bernard Cabane}
\email[]{bernard.cabane@espci.fr}
\affiliation{LCMD, CNRS UMR 8231, ESPCI, 10 rue Vauquelin, 75231 Paris Cedex 05, France}
\author{Joaquim Li}
\affiliation{Max Planck Institute for Dynamics and Self-Organization (MPIDS), 37077 G\"ottingen, Germany}
\author{Franck Artzner}
\affiliation{Institut de Physique, CNRS UMR 6626, Univ Rennes, 35042 Rennes, France }
\author{Robert Botet}
\affiliation{Physique des Solides, CNRS UMR 8502, Univ Paris-Sud, F-91405 Orsay, France}
\author{Christophe Labbez}
\affiliation{ICB, CNRS UMR 6303, Univ. Bourgogne Franche-Comt\'e, Dijon, France}
\author{Guillaume Bareigts}
\affiliation{ICB, CNRS UMR 6303, Univ. Bourgogne Franche-Comt\'e, Dijon, France}
\author{Michael Sztucki}
\affiliation{ESRF-The European Synchrotron, CS40220, 38043 Grenoble Cedex 9, France}
\author{Lucas Goehring}
\email[]{lucas.goehring@ds.mpg.de}
\affiliation{Max Planck Institute for Dynamics and Self-Organization (MPIDS), 37077 G\"ottingen, Germany}

\begin{abstract}
{We report small-angle x-ray scattering (SAXS) experiments on aqueous dispersions of colloidal silica with a broad monomodal size distribution (polydispersity 14\%, size $8$ nm).  Over a range of volume fractions the silica particles segregate to build first one, then two distinct sets of colloidal crystals.  These dispersions thus demonstrate fractional crystallization and multiple-phase (bcc, Laves AB$_2$, liquid) coexistence. Their remarkable ability to build complex crystal structures from a polydisperse population originates from the intermediate-range nature of interparticle forces, and suggests routes for designing self-assembling colloidal crystals from the bottom-up.}
\end{abstract}

\maketitle

\begin{figure}
\begin{center}
\includegraphics[width=85mm]{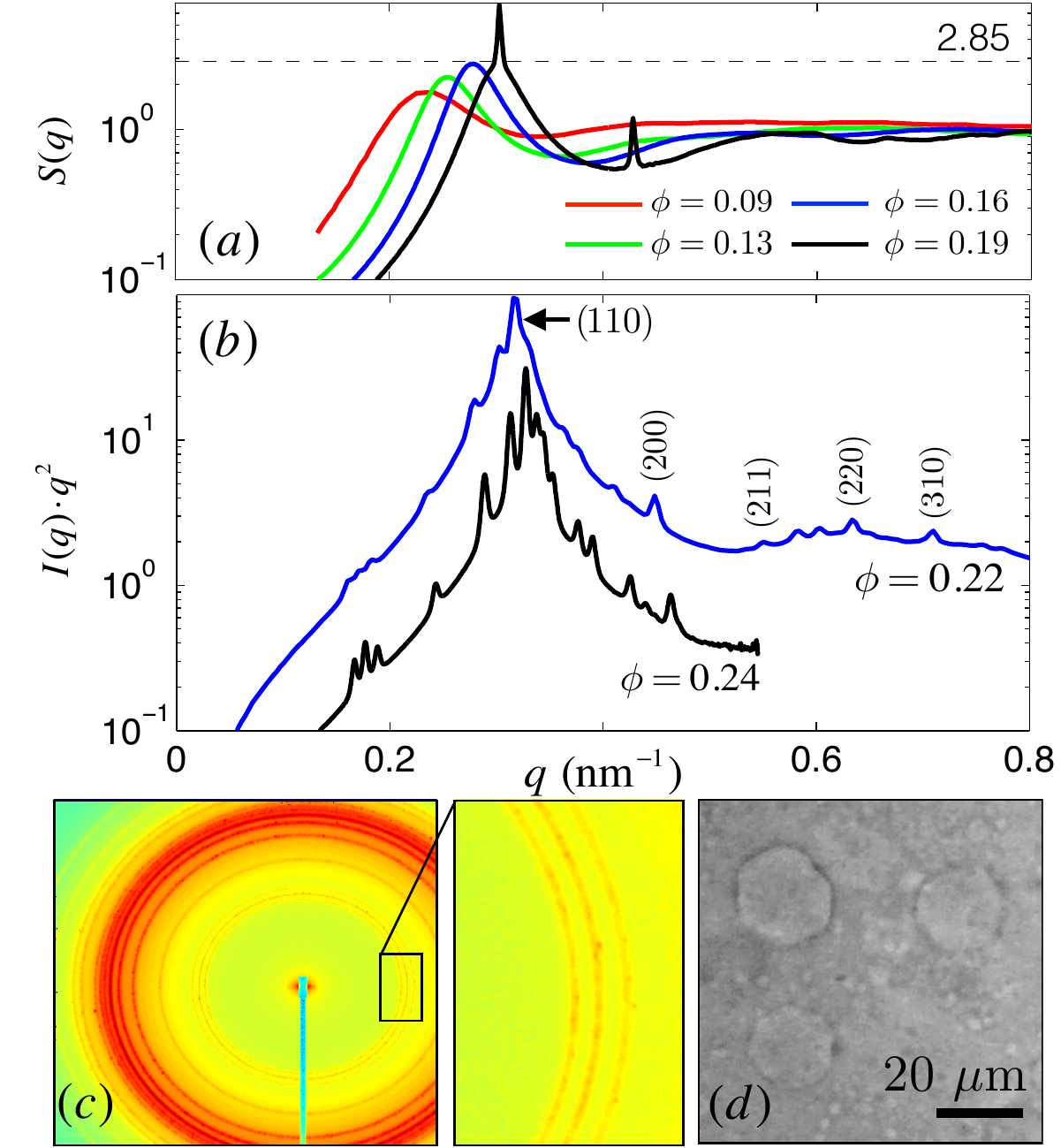}
\caption{(a) Effective structure factors. At low volume fractions, $\phi$, the dispersion has a liquid structure, with broad peaks. When the intensity of the liquid peak would exceeded 2.85, bcc colloidal crystals appear alongside the liquid phase.  (b)  At higher $\phi$ the scattering spectra show many sharp peaks in addition to the (indexed) bcc peaks.  Their positions and relative intensities correspond to crystals of a Laves MgZn$_2$ phase, in coexistence with the bcc and liquid phases.    (c) These diffraction patterns consisted of spots arranged in rings, and imply the existence of many micron-sized crystallites, which (d) can be seen directly by microscopy.  
\label{fig_struct}}
\end{center}
\end{figure}

What is the preferred structure for a population of colloidal particles, dispersed in liquid? This simple question has been satisfactorily answered only in the case of spherical particles that are effectively monodisperse in size \cite{Kose1973,Pusey1986,Pusey2009,Poon2015,Monovoukas1989,Russel1989}.  As the volume fraction of particles increases, there is a well-defined transition from a liquid to a crystal state. Two types of structures can be found, close-packed and body-centered cubic crystals; the preferred form depends on the range of interparticle forces~\cite{Monovoukas1989,Russel1989,Yethiraj2003}.

Polydisperse populations are a tougher problem. In one limit, for particles that interact as hard spheres, crystalline order is destroyed by even small amounts of polydispersity \cite{Pusey1987,Pusey2009,Poon2015,Williams2001,Schope2006b}.  Charged particles interact instead via soft potentials, and are more tolerant of polydispersity, especially where they have an \textit{effectively} narrow size distribution, due to long-range interactions.  In this other limit a crystal state can be retained at low volume fractions regardless of significant size polydispersity, if the interaction polydispersity remains low \cite{Leunissen2007,Russel1989,Lorenz2009}. Between these two limits is a vast region of phase space where we do not  know whether homogeneous crystallization or fractionated crystallization are possible.

Here we address the self-organization of polydisperse populations of particles that interact through forces with an \textit{intermediate} range, comparable to the variations in particle size. Using high-resolution scattering methods, we find that such populations can evolve through fractionated crystallization to yield coexisting crystals with different structures. These crystals can have large, complex unit cells with specific sites for particles of different sizes.  To explain this result, we use numerical simulations to demonstrate how a broad distribution of particles can split spontaneously into different types of crystals, which cooperate to make the best use of the whole population.

The colloids that we have used are industrially produced.  They consist of nanometric silica particles, dispersed in water (Ludox HS40).  The particles are roughly spherical with an average radius of 8 nm and a size polydispersity of 0.14 \cite{Goertz2009,Li2012}. We used near-equilibrium dialysis to equilibrate them against NaCl solutions (5 mM, pH 9.5). They were then slowly concentrated by addition of poly(ethylene glycol) to the solution outside the dialysis membranes, as in \cite{Jonsson2011,Li2012}. Under these conditions the particles repel each other via a screened electrostatic interaction, with a Debye length of 2.5-4.5 nm, depending on their volume fraction $\phi$.  Further details of our methods, and the dispersion properties (\textit{e.g.} charge, equation of state, density), are given as supplemental information \cite{SI}.

Samples were characterized through small-angle x-ray scattering (SAXS), using ID02 at ESRF.   The strength of ordering in a colloidal dispersion can be evaluated by the height, $S_{max}$, of the main peak of its effective structure factor $S(q)$, for scattering vector $q$ \cite{Verlet1968,Hansen1969,Beurten1981}.  $S(q)$ was found by dividing the radially-averaged scattering intensity $I(q)$ by the form factor of a dilute ($\phi = 10^{-3}$) dispersion, and normalizing at high $q$, as in \cite{DÕAguanno1991,Beurten1981,Pedersen1994,Li2012}.   For low $\phi$ these $S(q)$ had a broad main peak, indicative of disordered liquid arrangements of particles (Fig. 1a). Indeed, all these samples also behaved rheologically as fluids. The value of $S_{max}$ (Table \ref{tab_samples}) rose slowly with increasing $\phi$, from 1.2 at $\phi$ = 0.04, to 2.6$\pm0.1$ at $\phi$ = 0.16.   Despite our polydispersity, which should lower $S_{max}$ slightly \cite{DÕAguanno1991,Beurten1981}, and add a low-$q$ incoherent scattering \cite{DÕAguanno1991,Pedersen1994}, these values agree well with the Hayter-Penfold MSA model \cite{Hayter1981} of monodisperse Yukawa spheres (Table \ref{tab_samples}, using 8 nm particles with 5 mM salt and a surface charge of 170 $e$).

\begin{table}
\small
\centering
\begin{tabular}{c c c c}
\ \ $\phi \  \ $ \ & \ $S_{max}$ (liq.) & \ $S_{max}$ (MSA) \ & \ Phases \ \\ \hline
0.038	&1.2	& 1.33 & liquid\\
0.046	&1.4	& 1.40 & liquid\\
0.057	&1.5	& 1.50 & liquid\\
0.067	&1.6	& 1.58 & liquid\\
0.079	&2.2	& 1.69 & liquid\\
0.085	&1.8	& 1.74 & liquid\\
0.128	&2.2	& 2.12 & liquid\\
0.131	&2.1	& 2.15 & liquid\\
0.159	&2.7	& 2.42 & liquid\\
0.161	&2.5	& 2.44 & liquid\\
0.188	&-- & 2.72 & liquid, bcc\\
0.207	&--	& 2.94 & liquid, bcc\\
0.219	&--	& 3.08 & liquid, bcc, Laves\\
0.235	&--	& 3.28 & liquid, bcc, Laves\\
0.240	&--	& 3.35 & liquid, bcc, Laves \\ 
\end{tabular}
 \caption{Sample summary, showing the volume fraction $\phi$ ($\pm 0.005$), the intensity of the liquid peak $S_{max}$ and its predicted value (MSA) using \cite{Hayter1981}, and the observed phases.}
\label{tab_samples}
\end{table}

At $\phi$ = 0.19 and 0.21 we found that the 2D interference patterns of our dispersions also contained sharp diffraction spots, superimposed on the liquid-like scattering ring.  The spots are the powder-diffraction pattern of small crystallites.  Here any fractionation between the liquid and crystals would invalidate the decomposition of $I(q)$ into a form factor and effective structure factor.  Instead, we calculated the complex structure factor $F\sim I(q) \cdot q^2$, which does not require knowledge of the form factors of each phase. The positions of the peaks of $F(q)$, as well as systematic extinctions ($h+k+l $ odd), indicated that they originated from colloidal crystals with a body-centered cubic (bcc) structure.  This is in empirical agreement with liquid state theory, where, according to Verlet and Hansen \cite{Verlet1968,Hansen1969}, the liquid state with short-range order is unstable with respect to a crystalline structure when $S_{max} > 2.85$.  However, our dispersions were quite polydisperse, while the Verlet-Hansen criterion is strictly true only for monodisperse populations. Our observations suggest a possible reason why this agreement may still hold.  It involves growing the bcc crystals from a narrow subset (\textit{i.e.} an effectively monodisperse set) of the original population, and leaving the remaining particles in a liquid phase that coexists with these crystals. 

As the dispersions were compressed to higher $\phi$, between $0.22-0.24$, their scattering spectra became more complex. The interference patterns of these dispersions revealed a large number of spotty rings (Fig. 1c).    Typically hundreds of spots were seen, whose diameters, $\delta\simeq 0.003$ nm$^{-1}$, imply the presence of many crystallites with a diameter of at least $\pi/\delta = 1$ $\mu$m.  Microscope images (Fig. 1d) of such dispersions confirm the presence of stable free-floating crystals.

In these spectra we detected, after radial averaging, a broad liquid peak, peaks from the bcc phase, and up to 14 additional well-resolved peaks, including a triplet at low $q$, implying the presence of a crystal phase with a large unit cell. The new peaks can all be indexed (see Table \ref{tab_laves} and Supplemental Info \cite{SI}) to the powder spectrum of a crystalline phase of compact hexagonal (P6$_3$/mmc) symmetry, with lattice constants $a$ = 43.58 nm and $c = \sqrt{8/3}a$, and a unit cell volume of $\sqrt{2}a^3 = 1.17\times10^5$ nm$^3$, in the $\phi = 0.235$ sample.   In the same sample the bcc peaks were indexed to a unit cell with lattice constant $a_{bcc}$ = 27.11 nm and volume $a_{bcc}^3 = 1.99\times 10^4$ nm$^3$.  The unit cell volume of the new phase is therefore 5.9 times larger than that of the co-existing bcc phase, which contains 2 particle sites per cell.  Assuming that the number density of sites is comparable in both phases, which in conditions of close equilibrium and not too large fractionation is reasonable, one finds that the new phase has 12 particles per unit cell.

One can reasonably expect that this phase is constituted by a mixture of nanoparticles with distinct mean diameters. Among the varied options \cite{Filion2009} only one is of the compact hexagonal space group and contains 12 atoms per unit cell: the MgZn$_2$ Laves phase.  Here four Mg atoms are on the four equivalent $f$ Wyckoff positions, while eight Zn atoms are distributed on the six $h$ and two $a$ positions. This suggests that the new phase is composed of particles with two or three separate sizes organized into a Laves phase \cite{Berry1953}. Within this hypothesis, the intensities of the Bragg peaks were fit with three free parameters corresponding to the radii $r_a$, $r_f$ and $r_h$ of particles at the $a$, $f$, and $h$ sites (see supplementary information \cite{SI}). The fit, the results of which are shown in Table \ref{tab_laves}, converges when $r_f = 9.1\pm0.3$ nm and $r_a=r_h=7.3\pm0.3$ nm.  The stoichiometry is consequently AB$_2$ with four large particles and eight small particles per unit cell. The larger particles occupy relatively spacious truncated tetrahedron environments, where they are comfortably surrounded by rings of smaller particles in octahedral sites (Fig.~\ref{fig_struct}).  Irast to repulsive monodisperse crystals \cite{Pusey1986}, the density of this Laves phase thus appears to be slightly lower (0.22) than that of the coexisting liquid (0.235); this situation could relate to the size selection of the individual sites.

\begin{figure}
\begin{center}
\includegraphics[width=75mm]{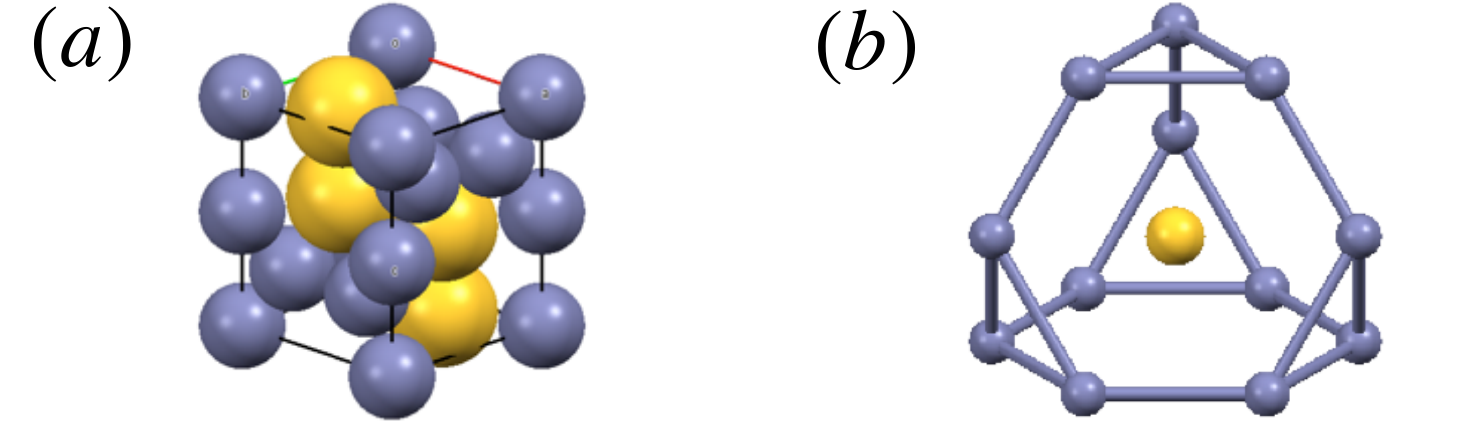}
\caption{(a) Unit cell of the Laves MgZn$_2$ phase.  (b) Larger-than-average particles (yellow) occupy central sites, and are surrounded by rings of smaller-than-average particles (blue).  
\label{fig_struct}}
\end{center}
\end{figure}

Various AB$_2$ phases are well-known in binary mixtures of hard spheres \cite{Bartlett1992,Schofield2005,Sanders1980,Murray1980,Shevchenko2005,Hynninen2007}. For example, the AlB$_2$ structure is a preferred crystal phase for binary mixtures with a size ratio of the smaller to the larger particles between about 0.4 and 0.6 \cite{Schofield2005} and occurs in gem opals \cite{Sanders1980,Murray1980}, while the MgCu$_2$ phase can be templated by walls \cite{Hynninen2007}. What we have shown, however, is that similar phases also naturally arise in the solidification of broad and continuous populations of nanoparticles. 

\begin{table}
\small
\centering
\begin{tabular}{c c c c c c}
\ $h \	k \ l$ \ & \ $m$ \ &\ $q_{exp}$ (nm$^{-1}$)\ &\ $q_{fit}$ (nm$^{-1}$)\ &\ $F_{exp}$\ &\ $F_{fit}$\ \\   \hline  
0	0	1&2	&not obs.	& 0.0883	&not obs.	& 0 \\
1	0	0&6&	0.1667&	0.1665&	12.9&	12.9 \\
0	0	2&2&	0.1769&	0.1766&	26.9&	21.6\\
1	0	1&12&	0.1885&	0.1884&	9.0&	12.0\\
1	0	2&12&	0.2431&	0.2427&	18.1&	21.9\\ 
0	0	3&2&	not obs.&	0.2648&	not obs.&	0\\
1	1	0&6&	0.2891&	0.2883&	84.3&	84.3\\
1	1	1&12&	not obs.&	0.3015&	not obs.&	0\\
1	0	3&12&	0.3132&	0.3128&	98.2&	79.4\\
2	0	0&6&	0.3329&	0.3329&	52.4&	42.3\\ 
1	1	2&12&	0.3378&	0.3381&	73.7&	87.6\\
2	0	1&12&	0.3441&	0.3444&	57.5&	76.2\\
0	0	4&2&	0.3530&	0.3531&	86.2&	84.8\\
2	0	2&12&	0.3767&	0.3768&	29.1&	32.1\\
1	0	4&12&	0.3903&	0.3904&	25.8&	25.8\\ 
1	1	3&12&	not obs.&	0.3915&	not obs.&	0\\
2	0	3&12&	0.4256&	0.4254&	19.6&	18.2\\
2	1	0&12&	0.4402&	0.4404&	8.4&	10.0\\
\end{tabular}
 \caption{Positions and relative scattering intensities of the observed and fitted diffraction peaks of the Laves phase, for $\phi = 0.235$. $F$ is a complex structure factor corrected for the multiplicity of the peaks, $m$, and the averaging of the powder diffraction pattern; zero indicates a systematic extinction.}
\label{tab_laves}
\end{table}

An explanation for the coexistence of different crystal types, each composed of a subset of particle radii, can be made by seeking the equilibrium phases of the particle population. To this end we investigated the fractionation of polydisperse charged particles through Gibbs-ensemble Monte-Carlo numerical simulations \cite{Panagiotopoulos1988} of a combination of a Laves MgZn$_2$ phase and a bcc phase, with an fcc phase added as a control.  The model is similar to that in \cite{Botet2016}. Each phase was treated as an isolated volume (avoiding grain boundaries), but particles could move randomly between sites within each phase, and between phases, according to a Monte-Carlo Metropolis algorithm at room temperature \cite{Landau2000}.  Although, for simplicity of demonstration, no colloidal liquid was modeled, we would expect such a phase to act as the medium of particle exchange, and an acceptor of misfit particles.  The proportions of particles and the lattice constants of the three phases were allowed to vary with volume exchange between them, keeping the total volume constant. 

We considered a model of 22466 particles with a Gaussian distribution of sizes $r$, an average radius of 8 nm and a polydispersity of 0.14 \cite{Goertz2009}, with a global $\phi=0.22$.  Interactions between particles were modeled as hard core plus Yukawa pair-potentials, with an effective Debye length of $\kappa^{-1} = 2.8$ nm and effective surface charge density of 0.2 $e$/nm$^2$ (\textit{i.e.} the charge on particle $i$ scales as $r_i^2$).  These parameters are estimated as in \cite{Alexander1984,Belloni1998,Trizac2003}, accounting for modest charge renormalization, and agree with the dispersion's experimentally determined equation of state \cite{Jonsson2011,Li2015}. 

\begin{figure}
\begin{center}
\includegraphics[width=85mm]{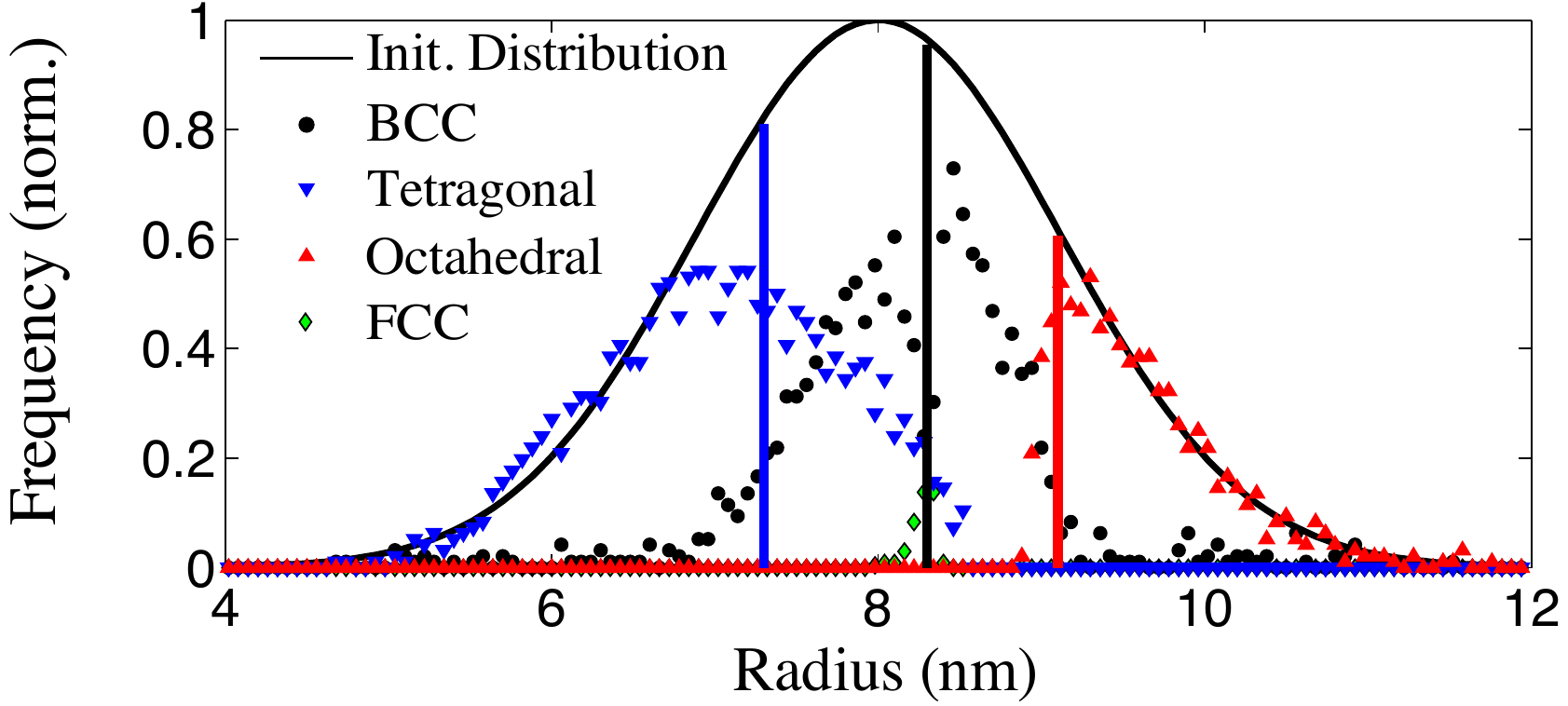}
\caption{Monte Carlo simulations of the fractionation of a polydisperse colloidal dispersion into three preset crystalline structures (bcc, fcc, Laves MgZn$_2$). Shown are the final equilibrium particle-size distributions in each phase.   The vertical lines show the average radii extracted from the SAXS data for particles in the bcc phase (black), and in the tetragonal (blue) and octahedral (red) sites of the Laves phase. 
\label{fig_botet}}
\end{center}
\end{figure}

Over time, the system evolved to find a configuration of minimal Madelung energy, and the proportion of each phase stabilized; Fig. \ref{fig_botet} shows the final distribution of particle sizes, according to phases and sites.    It shows how the coexistence of a Laves phase with the bcc phase is possible: the bcc phase uses the most populated part of the distribution of particle sizes, near the centre of the distribution. In this example a small minority of particles was also taken into the fcc phase, although this phase disappears if a longer screening length (3 nm) is used.  In either case, the remaining particles have a bimodal size distribution and thus fit efficiently into the differently shaped sites of the Laves phase.   Exploring different parameter values, we found that these results were robust to doubling the charge density of the particles, or letting their charge scale with $r$ (as occurs for strong charge condensation \cite{Gabrow2004}), but sensitive to changes to the effective screening length (between 2.2 and 3.0 nm).  The model's average radii of 7.0, 8.2, and 9.6 nm, for particles at equilibrium in the Laves tetragonal sites, bcc sites, and Laves octahedral sites, respectively, correspond well to the corresponding experimental values of 7.3, 8.3, and 9.1 nm.  

We have thus described how a polydisperse population can split into coexisting phases of a colloidal liquid, a bcc crystal that preferentially selects the most abundant particle sizes, and a Laves phase that accommodates the remaining bimodal distribution of particles. This segregation by particle size is known as fractionated crystallization; similar processes are known in molecular systems \cite{Timms2005}, including geochemistry \cite{Ghiorso1985}.  For hard-sphere colloids fractionation has been predicted beyond a terminal polydispersity of about 6\% \cite{Bartlett1998,Bartlett2000,Fasolo2003,Sollich2010,Bolhuis1996}.   For medium-range Yukawa interactions ($\kappa a$ between 2.5-10), recent simulations \cite{vanderLinden2013} have suggested that a size polydispersity of 10-15\%, comparable to ours, is required to hinder crystallization, and thus potentially trigger fractionation. 

Experimentally the best prior evidence of colloidal fractionation is the work of van Megen and collaborators \cite{Martin2003,Schope2006,Schope2006b}, who invoke it to explain the nucleation processes of colloidal crystals near a terminal polydispersity.   The coexistence of multiple solid phases is also known in cases of low-dimensional systems such as platelets \cite{Byelov2010} or particles confined to a plane \cite{Eiser2010}. Further evidence may also be hiding in old data such as Fig. 13 of Ref. \cite{Chang1995}, which appears to imply the presence of large-unit-cell crystals in dispersions similar to ours (10.2 nm silica with 9\% size polydispersity).  

The fractionation of particles in our experiments depends on their intermediate range of interactions.  Much work on colloidal crystals is performed with particles that interact as hard spheres, and which crystallize when they are in close to direct contact, at $\phi\sim0.5$. When such particles have a broad distribution of sizes, then the unavoidable overlaps of any large adjacent particles inhibit the formation of a structure with long-range order  \cite{Pusey1987,Pusey2009,Bolhuis1996,Poon2015}, and dynamic arrest turns the dispersion into a glass \cite{Pusey1987,Pusey1986}. Our particles interact instead through soft potentials. Assuming an effective Yukawa potential  \cite{Alexander1984,Belloni1998,Trizac2003}, the pair-potential of two average-sized particles reaches about 3 kT at a volume fraction of 20\%, corresponding to a surface separation (for bcc) of 8 nm.   In this state, overlap of the particles themselves is still a rare occurrence, determined by the frequency of very large particles. These few ``outliers" can easily be rejected away from the surfaces of growing crystals, as the soft potentials also keep the mobility of such particles high.  

The width of the particle size distribution and the range of particle interactions together control the frequency of such outliers, which are then available to build more diverse structures. We consider three cases.  If the interactions are long range (effective diameter $\gg a$), then variations in the particle size will be screened, and simple fcc or bcc crystals are both expected and seen \cite{Kose1973,Leunissen2007,Russel1989,Lorenz2009}.  If the interaction range is intermediate, for example $\kappa a \sim 1$, but the polydispersity $\sigma$ is too high, then there will be too many overlaps to nucleate the first bcc crystals, and the dispersion may remain in a liquid or glass phase.  Inverting Pusey's criterion \cite{Pusey1987} suggests that this will be the case when $\phi \geq c(1/(1+\sigma))^3$, where the order-1 constant $c$ depends on how tolerant a crystal is to overlaps.  If, however, the effects of the soft potential and the number of overlaps are balanced against each other, as in this letter, then fractionation is encouraged, and the phase space of polydisperse colloidal dispersions is opened.

The behavior of such polydisperse nanometric dispersions points to directions that have not been explored so far, despite theoretical predictions \cite{Bolhuis1996,Fasolo2003,Fernandez2007,Bartlett1998,Sollich2010,vanderLinden2013}.  We demonstrate here fractionated crystallization, with coexistence of at least three very different phases (liquid, bcc and Laves), and the formation of complex crystals that efficiently utilize the full size distribution. The link between the particle size distribution and the structures also gives us a scheme for generating even more complex phases through the crystallization of populations of particles with broader size distributions, provided that they interact through soft medium-range potentials.  The variety of structures waiting to be discovered could be enormous, given that, within the limits defined above, there exists a huge phase space of different size distributions and interaction potentials to explore.

\newpage
%\onecolumngrid

        \setcounter{table}{0}
        \renewcommand{\thetable}{S\arabic{table}}%
        \setcounter{figure}{0}
        \renewcommand{\thefigure}{S\arabic{figure}}%
           \setcounter{figure}{0}
        \renewcommand{\thesection}{S\arabic{section}}%

\setstretch{1.08}
\onecolumngrid
\section{Hiding in plain view:Colloidal self-assembly from polydisperse populations\\ Supplemental Information}
\twocolumngrid
\section{Materials and methods}

Colloidal silica (Ludox HS40, Sigma-Aldrich) was cleaned and concentrated by the osmotic stress method, as detailed in \cite{Jonsson2011,Li2012,Li2015}.  Millipore (Milli-Q) deionized water was used for all steps. The surfaces of the silica particles were cleaned through prolonged exchange with an aqueous salt solution (NaCl 5 mM) at a controlled pH (all solutions measured between pH 8.8-9.5), across a dialysis membrane with a molecular cutoff of 14 kD.  The concentrations of ions in the dispersion are thus in Donnan equilibrium with NaCl at 5 mM.  The volume fractions of the colloids were then adjusted by the addition of poly(ethylene glycol) (PEG 35000, Sigma) to the solution outside the dialysis membrane \cite{Li2015}.  The surfaces of the particles were not treated in any other way, although we emphasize that this ``washing" process is important in order to obtain reliable results with particles that have exchangeable counter-ions.  After dialysis, samples were poured into Falcon tubes, sealed, and stored until use. 

We determined the volume fractions $\phi$ of our samples by weight measurements, before and after drying the dispersions overnight at 120-140$^\circ$C to eliminate adsorbed water.  Results were statistically reproducible to within $0.5\%$. To calculate $\phi$ we assumed a mass density of the silica particles of  2200 kg/m$^3$. This is consistent with the relation of the position of the liquid SAXS peak to silica volume fraction \cite{Li2012}, with contrast matching experiments for the same particles in D$_2$O + H$_2$O mixtures in SANS \cite{Wong1988}, with the manufacturerÕs specifications, and numerous previous publications using similar dispersions (see \textit{e.g.} \cite{Roberts2005,Iler1979}).  Allowing for up to an error in density of $\pm$50 kg/m$^3$ would introduce a systematic error into the $\phi$ measurements of no more than 0.3\%.  Note that the density of Ludox particles is close to that of amorphous silica, in contrast to the lighter micro-porous particles that are instead synthesized by the St\"ober process \cite{vanBlaaderen1993,Masalov2011}.

The experiments described in this study were performed over four SAXS sessions, using three separate series of dialysis, with different stock bottles each time.  All experiments were conducted using the instrument ID02 at ESRF at a fixed wavelength of 0.1 nm (12.4 keV) with a spread in wavelength of $\leq$0.015\%.  An elliptical beam was used in all cases, with a height (full-width-half-maximum) of 50-70 $\mu$m, and a width of  250-400 $\mu$m \cite{Li2012,Boulogne2014}, and with divergences of 20 $\mu$rad and 40 $\mu$rad, respectively.  Spectra were collected at detector distances of 1 m, 2.5 m, and 10 m. In all cases the beam was centered on the middle of the sample cell, and the photon fluxes used were of order $5\cdot10^{12}$ s$^{-1}$. 

Three different types of cells were used: quartz glass capillary tubes (Hilgenberg) with an inner diameter of 1.3 mm, a length of $\sim$8 cm and wall thickness of 0.01 mm; standard steel cells from the beam line, with mica windows (Richard Jahre GmbH, 10-20 $\mu$m thickness), an 8 mm inner diameter and a path length of 0.5 mm; and single-use cells made from trapping a drop of dispersion (transferred to the cell by pipette) between two kapton films, separated by a $\sim$0.5 mm flexible ring.  The capillaries were inserted into the capillary sample changer of ID02, translated sequentially to a position intersecting the beam, and exposed to the beam for very short times (0.1 to 1 s). A similar procedure was applied for the steel and kapton cells.  In all cases the backgrounds spectra of empty cells were subtracted from the scattering spectra before further processing. Microscope observations of the crystals were also made in 50 $\mu$m thick Hele-Shaw cells made from two standard microscope glass slides.  Finally, we note that samples were not subject to any shear-melting regime prior to use.  Instead, all samples started as a colloidal liquid, and were concentrated over a period of weeks in the absence of bulk flow.   We found no effect of the type of cells, or the different preparations, on the phases observed, or the crystallization phenomena. 

The properties of these dispersions have been well-studied in the past, and we provide a summary here.  The full particle size distribution of Ludox HS40 has been measured directly through transmission electron microscopy \cite{Goertz2009}.  The particles are roughly spherical, with a mean radius of 8.15 nm and their distribution of radii is well-fit by a Gaussian with a polydispersity of 0.14.  SAXS measurements on one of the samples used in our experiments \cite{Li2012} confirm these values: the form factor of a dilute dispersion was consistent with a mean diameter of 8.0 nm and a size polydispersity of 0.14.  

\begin{figure}
\begin{center}
\includegraphics[width=85mm]{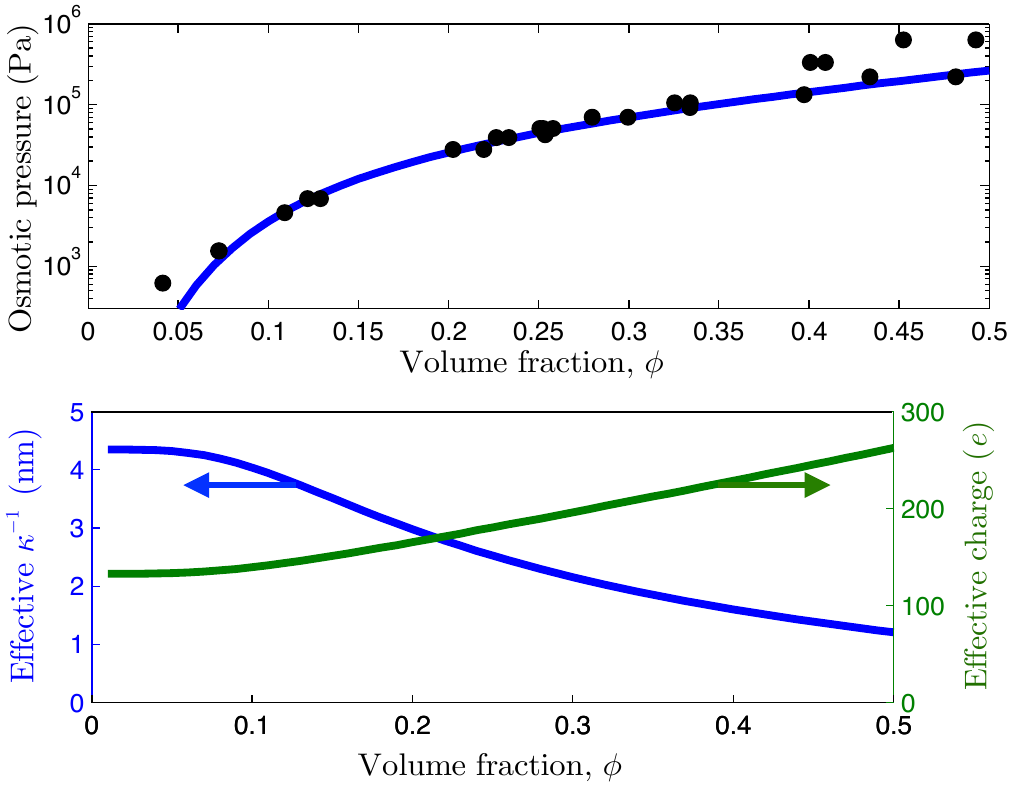}
\caption{(a) The osmotic pressure of Ludox HS40, dialyzed against 5 mM NaCl at pH 9, was measured in Ref. \cite{Jonsson2011}.  Shown here are their data (black points), accounting for the corrected equation of state for PEG 35000 from Ref. \cite{Li2015}, along with the predictions (blue) of a Poisson-Boltzmann cell model.  (b) The effective Debye length ($\kappa^{-1}$, blue curve) and charge per particle (green), can be calculated by the same model, for average-sized (8 nm) particles, with a bare charge of 402 $e$ (\textit{i.e.} a surface charge density of 0.5 $e$/nm$^2$).   
\label{fig_pressure}}
\end{center}
\end{figure}

The bare, or surface, charge of silica nanoparticles has been measured for various dispersions at different ionic strengths and pH values \cite{Bolt1957,Persello2000}.  For 8 nm Ludox dialyzed against 1 mM NaCl, Bolt \cite{Bolt1957} reports a surface charge density in the range of 0.3-0.5 $e$/nm$^2$, between pH 9-10.  Persello \cite{Persello2000} gives a slightly higher value of 0.6 $e$/nm$^2$ for a colloidal silica (S 22) at 5 mM NaCl, pH 9.  Finally, we note that a Poisson-Boltzmann cell model using a bare charge of 0.5 $e$/nm$^2$ fits the experimental osmotic compression curves for different colloidal silicas at pH 9 and a range of salt concentrations \cite{Jonsson2011,Li2015}.  We adopt this value here, and note that as it scales as the radius squared, the polydispersity of the bare charge is about twice that of the radius, or 0.28.

Due to charge condensation (see \textit{e.g.} \cite{Alexander1984,Belloni1998,Trizac2003}), the effective interactions of our particles are related to a reduced, or renormalized charge.  To estimate these effects we used the Poisson-Boltzmann cell model in the form summarized by Belloni \cite{Belloni1998}.  This model solves the non-linear Poisson-Boltzmann equation on an electrically neutral spherical cell surrounding each colloidal particle.  In \cite{Jonsson2011} this model was shown to match both a more detailed Monte Carlo simulation of the ionic distributions around silica nanoparticles, and the observed osmotic pressures of Ludox HS40 under our experimental conditions [see Fig. \ref{fig_pressure}(a)]. From it we calculated the effective interactions of particles in our dispersions using Alexander's prescription \cite{Alexander1984,Trizac2003}.  Specifically, we used Eqs. 6 and 16 of Ref. \cite{Trizac2003} to calculate the effective interaction length and effective charge for a Yukawa potential between two average-sized (8 nm, surface charge 402 $e$) particles, at various concentrations, as shown in Fig. \ref{fig_pressure}(b).  The values at $\phi = 0.22$ ($\kappa^{-1} = 2.8$ nm, effective charge 171 $e$) were used as inputs to the Monte-Carlo model described in our letter.   

Finally, within the cell-model we also investigated the effects of changing the particle radius on the reduced charge, in an attempt to evaluate the charge and interaction polydispersity.  For strong charge renormalisation \cite{Gabrow2004} it is known that the reduced charge scales linearly with the average particle radius $a$.  In the absence of charge renormalisation, it should scale as the bare charge, namely $a^2$.  We found that, for our small particles at intermediate salt concentrations, the reduced charge scales in an intermediate way, of approximately $a^{1.4}$, for small changes around $a = 8$ nm.  Converting this into a charge polydispersity would allow us to estimate a reduced charge polydispersity of 19\%, arising from the particle size polydispersity.  Since the Yukawa potential is a pair potential that scales with the independent charges on two particles, the interaction polydispersity of the effective potential is $\sqrt{2}$ times higher than that of the charge polydispersity.  

\section{Powder diffraction analysis}

For a powder diffraction pattern the intensity $I$ of a Bragg peak with Miller indices $hkl$ at a scattering vector $q$ is 
\begin{equation} \label{intensity}
I(hkl) = \frac{|F(hkl)|^2 m(hkl)}{q^2} e^{-q^2\left\langle u^2 \right\rangle/3}.
\end{equation}
Here $F(hkl)$ is the complex structure factor of the unit cell and $m(hkl)$ is the multiplicity of the peaks.  The exponential term is the Debye-Waller factor, which accounts for thermal fluctuations of particles around their equilibrium positions: $\left\langle u^2 \right\rangle$ is the mean squared displacement induced by thermal agitation.  Finally, the 1/$q^2$ correction is due to the spreading of the Bragg peak in reciprocal space, over a sphere of radius $q$.    Note that the definition of the complex structure factor $F$ is different from that of the effective structure factor $S$, which is discussed in our letter with respect to $liquid$-like structures.  In particular, $F$ can be measured without making any assumptions about the fractionation of particles into any individual co-existing phase.

The complex structure factor can be found by summing over the contributions of all objects in a unit cell
\begin{equation} \label{F_theory}
F(hkl) = \sum_nf_nA_n
\end{equation}
where $A_n$ is a geometrical factor related to the arrangement of the objects, and $f_n$ relates to the shape of the individual scattering objects.  For the case of monodisperse spherical nanoparticles of radius $r$, 
\begin{equation}
f_n(q,r) = \frac{4\pi r^3}{(qr)^3} (\sin(qr)-qr\cos(qr)).
\end{equation}
Noteworthy, the polydispersity of the particles occupying each site does not have any effect on the relative intensities, because any independent form factor fluctuations result in a $q$-constant increase of the SAXS background. Only spatially correlated form factor fluctuations should cause both the Bragg peak intensity to decrease and additional diffuse scattering, but this is not observed in the present case

\subsection{Structural Analysis of Laves phase}

For volume fractions $\phi$ = 0.219, 0.235, and 0.240, we found up to 14 peaks of $I(q)$ corresponding to colloidal crystals arranged as a MgZn$_2$ Laves phase.  The scattering spectrum of the $\phi=0.235$ sample was of slightly better quality, and its analysis is presented here (the other spectra are consistent with the same structure).    The position, width, and height of each peak was fit using a Lorentzian line-shape, allowing for a slowly varying background.  The half-width-half-maxima, $\delta$, of all these peaks were approximately equal, and between 0.003-0.004 nm$^{-1}$ (compared to an instrument resolution of 3$\cdot$10$^{-4}$ nm$^{-1}$). This indicates the absence of any disorder of the second kind (long-range) in the crystals and demonstrates their high positional quality. The constant width of the peaks shows that the crystals are at least of a size $\pi/\delta$, or 1 micron.  Thus, the crystals must be at least of order a hundred particles across.

The positions of the observed peaks can all be indexed to the reflections of the hexagonal crystal system.  For this system, scattering peaks are possible when
\begin{equation}
q = 2\pi\bigg(\frac{4}{3}\big(\frac{h^2+hk+k^2}{a^2}\big) + \frac{l^2}{c^2}\bigg)^{1/2}.
\end{equation}
Table I (main text) compares the positions of the observed and predicted scattering peaks for fitted lattice constants $a = 43.58$~nm and $c  = 71.17$~nm $ = \sqrt{8/3}a$.   The point group must have the highest symmetry because of the spherical symmetry of the particles, \textit{e.g.} 6/mmm. However, the high quality of the data shows clearly the extinction of the (0,0,1), (0,0,3), (1,1,1), and (1,1,3) reflections, indicating a glide-mirror along $c$.  The space group is consequently compact hexagonal (No. 194, P6$_3$/mmc).  The unit cell has a volume of $V_0 = \sqrt{2}a^3 = 117 050$~nm$^3$, or 11.7 times the volume occupied by a nanoparticle in the coexisting bcc phase in the same sample (see analysis in Sect. \ref{sect_bcc}).

The colloidal MgZn$_2$ Laves phase is constituted by 4 large nanoparticles and 8 small nanoparticles, arranged within a unit cell of the compact hexagonal space group.  As measured relative to the edges of the unit cell, the large particles are at coordinates
\begin{equation}
(x,y,z) = \bigg\{(\frac{1}{3},\frac{2}{3},\frac{1}{16}), (\frac{1}{3},\frac{2}{3},\frac{7}{16}), (\frac{2}{3},\frac{1}{3},\frac{9}{16}), (\frac{2}{3},\frac{1}{3},\frac{15}{16})\bigg\}  \nonumber
\end{equation}
whereas the small particles are at coordinates
\begin{align}
 (x,y,z) = \bigg\{(0,0,0), (0,0,\frac{1}{2}),(-\frac{1}{6},\frac{1}{6},\frac{1}{4}), (-\frac{1}{6},-\frac{1}{3},\frac{1}{4}), \nonumber\\ ... (\frac{1}{3},\frac{1}{6},\frac{1}{4}), (\frac{1}{6},-\frac{1}{6},\frac{3}{4}), (\frac{1}{6},\frac{1}{3},\frac{3}{4}), (-\frac{1}{3},-\frac{1}{6},\frac{3}{4})\bigg\}. \nonumber
 \end{align}
For the small particles, the first two coordinates correspond to the Wyckoff $a$ positions, while the last six coordinates are at the Wyckoff $h$ positions.  The large particles occupy the Wyckoff $f$ positions.   In this configuration, the geometric factor for each nanoparticle $n$ is 
\begin{align}
 A_n &= 8 \cos\big(2\pi[lz+l/4]\big) \\ \times  \big\{&\cos\big(\pi i [x+y]\big)\cos\big(\pi[(h-k)(x-y)-l/2]\big)\nonumber \\ + &\cos\big(\pi h [x+y]\big)\cos\big(\pi[(k-i)(x-y)-l/2]\big) \nonumber \\ +
 &\cos\big(\pi k [x+y]\big)\cos\big(\pi[(i-h)(x-y)-l/2]\big)\big\}\nonumber
 \end{align}
 where $h+k+i=0$.
 
We converted the experimental scattering intensities into $F_{exp}$, and compared them with the calculated complex structure factors $F_{fit}$ for an MgZn$_2$ lattice.  The peak intensities are well-fit with only three free parameters, the radius of the small particles $r_s = 7.3\pm0.3$ nm, the radius of the large particles $r_l = 9.1\pm0.3$ nm, and the amplitude of the thermal fluctuations $\left\langle u^2 \right\rangle = (1.8$ nm$)^2$.  If we further allow the radii of the smaller particles at the $a$ and $h$ Wyckoff positions to vary independently, we find that they both converge to the same $r_s$.  
 
\subsection{Structural Analysis of bcc phase} \label{sect_bcc}

\begin{table}
\small
\centering
\begin{tabular}{c c c c c c}
\hline
\ $h \	k \ l$ \ & \ $m$ \ &\ $q_{exp}$ (nm$^{-1}$)\ &\ $q_{fit}$ (nm$^{-1}$)\ &\ $F_{exp}$\ &\ $F_{fit}$\ \\   \hline  
1	1	0&12	& 	0.303& 	0.303	&99	& 100 \\
2	0	0&6&	0.428&	0.428&	20.5&	22 \\
2	1	1&24&	0.524&	0.524&	-2.7&	-2.3\\
2	2	0&12&	0.606&	0.605&	-10&	-7\\
3	1	0&24&	0.677&	0.677&	-6.5&	-5.7\\ \hline
\end{tabular}
 \caption{Positions $q$ and magnitudes of the peaks of the complex structure factor $F$ for the observed and fitted diffraction peaks of the bcc phase in coexistence with the colloidal liquid, at $\phi = 0.188$.  The fit converged when $a = 29.35$~nm, $r_{bcc} = 8.8 \pm 0.3$~nm, and $\left\langle u^2 \right\rangle = (2.2$~nm$)^2$.}
\label{tab_bcc1}
\end{table}

\begin{table}
\small
\centering
\begin{tabular}{c c c c c c}
\hline
\ $h \	k \ l$ \ & \ $m$ \ &\ $q_{exp}$ (nm$^{-1}$)\ &\ $q_{fit}$ (nm$^{-1}$)\ &\ $F_{exp}$\ &\ $F_{fit}$\ \\   \hline  
1	1	0&12	& 	0.317& 	0.317	&100	& 100 \\
2	0	0&6&	0.449&	0.449&	25&	26 \\
2	1	1&24&	0.550&	0.550&	0&	-1\\
2	2	0&12&	0.634&	0.635&	-11&	-8\\
3	1	0&24&	0.711&	0.710&	-6&	-7\\ \hline
\end{tabular}
 \caption{Positions $q$ and magnitudes of the peaks of the complex structure factor $F$ for the observed and fitted  diffraction peaks of the bcc phase in coexistence with the colloidal liquid and Laves phase, at $\phi = 0.219$.  The fit converged when $a = 27.99$~nm, $r_{bcc} = 8.3\pm 0.3$~nm, and $\left\langle u^2 \right\rangle = (1.9$~nm$)^2$}
\label{tab_bcc2}
\end{table}

\begin{table}[t!]
\small
\centering
\begin{tabular}{c | c   c | c  c}
\hline
 \multicolumn{1}{c}{Parameter}  & \multicolumn{2}{c}{Experiment} &\multicolumn{2}{c}{Monte-Carlo} \\   \hline  
\  \ & \ Laves phase \ & \  bcc \ & \ Laves phase \ & \  bcc \ \\ \hline
lattice const. $a$ (nm) & 43.58 & 27.99 & 43.8 & 27.8 \\ \hline
particle radii (nm) &  $r_s = 7.3$  & $ 8.3$ & $r_s = 7.0$  & $ 8.2$ \\
\ & $r_l = 9.1$ & \ & $r_l = 9.6$ \\
\ & $\left\langle r \right\rangle = 7.9$ & \ & $\left\langle r \right\rangle = 8.2$ \\ \hline
inter-particle  &  $d_{s-s}$ = 7.3 & $ 7.6$ & $d_{s-s}$ = 7.5 & $ 7.6$ \\
distance (nm) \ &  $d_{s-l}$ = 8.5 & \ & $d_{s-l}$ = 8.2 & \ \\
  \ &  $d_{l-l}$ = 9.1 & \ & $d_{l-l}$ = 9.2 & \ \\ \hline
$\phi$ in crystal & 	0.217 & 0.218 & 0.22 & 0.22 \\ \hline
average (bulk) $\phi$	& 0.235 & 0.219 & -- &  -- \\
 \hline
\end{tabular}
 \caption{Summary of structural analyses, and a comparison between experimental observations and Monte-Carlo simulations.}
\label{tab_summary}
\end{table}

For the bcc phase, the geometrical factor of each particle (one at the origin of the unit cell, the other at its centre), is $A = 1$, if $h+k+l$ is even, and 0 otherwise.  Scattering peaks from bcc crystals (space group 229) are allowed at   
\begin{equation}
q = 2\pi \bigg(\frac{h^2+k^2+l^2}{a^2}\bigg)^{1/2}
\end{equation}
when $h+k+l$ is even, and $a$ is the lattice constant. 

The bcc peaks of several spectra were analyzed in detail. In each case, as with the Laves phase discussed above, the lattice constant $a$ was fit to the peak positions, while the average radius, $r_{bcc}$, of the particles in the bcc phase, and the thermal fluctuation amplitude $\left\langle u^2 \right\rangle$ were fit to match the distribution of peak intensities.  The results of the fits for $\phi = 0.188$ and $\phi = 0.219$ are shown in Tables \ref{tab_bcc1} and \ref{tab_bcc2}, respectively.  For the further situation $\phi = 0.235$, only the first two bcc peaks were visible, from which we could derive the lattice constant $a = 27.11$ nm.  The bcc unit cell contains 2 nanoparticles, and has a volume of $a^3$, giving a volume per particle of 9960 nm$^3$ for the $\phi = 0.235$ sample.

\subsection{Summary and comparison to Monte-Carlo simulation}

A summary of the structural analyses for the Laves and bcc phases is presented in Table \ref{tab_summary}, which also gives some geometrical parameters of both phases, and shows equivalent measurements from the Monte-Carlo simulation.  Briefly, in a bcc crystal of lattice constant $a$, the distance between the centers of adjacent particles is $\sqrt{3}a/2$.  However, the particles are not in contact, and the average separation of their surfaces is $d_{bcc} = \sqrt{3}a/2 - 2r_{bcc}$, where $r_{bcc}$ is the mean radius of the particles in the bcc phase.  In the case of the Laves phase, the surface-separations of adjacent small particles is $d_{s-s} = a/2 - 2r_s$, of adjacent large particles is $d_{l-l} = \sqrt{3/8}a - 2r_l$, and of adjacent large and small particles is $d_{s-l} = \sqrt{11/32}a - r_s - r_l$.  In all cases the nanoparticles are not in contact, and are separated by approximately the same gaps.  For the Monte-Carlo simulation, all values represent averages over all particles in a phase.

%\bibliography{crystalbib}


\end{document}